# A Hebbian/Anti-Hebbian Network for Online Sparse Dictionary Learning Derived from Symmetric Matrix Factorization

Tao Hu[1], Cengiz Pehlevan[2,3], and Dmitri B. Chklovskii[3]

[1] Texas A&M University
MS 3128 TAMUS
College Station, TX 77843
taohu@tees.tamus.edu

[2] Janelia Farm Research Campus
Howard Hughes Medical Institute
Ashburn, VA 20147
pehlevanc@janelia.hhmi.org

[3] Simons Center for Data Analysis
Simons Foundation
New York, NY 10010
mitya@simonsfoundation.org

*Abstract*—Olshausen and Field (OF) proposed that neural computations in the primary visual cortex (V1) can be partially modelled by sparse dictionary learning. By minimizing the regularized representation error they derived an online algorithm, which learns Gabor-filter receptive fields from a natural image ensemble in agreement with physiological experiments. Whereas the OF algorithm can be mapped onto the dynamics and synaptic plasticity in a single-layer neural network, the derived learning rule is nonlocal - the synaptic weight update depends on the activity of neurons other than just pre- and postsynaptic ones – and hence biologically implausible. Here, to overcome this problem, we derive sparse dictionary learning from a novel cost-function - a regularized error of the symmetric factorization of the input's similarity matrix. Our algorithm maps onto a neural network of the same architecture as OF but using only biologically plausible local learning rules. When trained on natural images our network learns Gabor-filter receptive fields and reproduces the correlation among synaptic weights hard-wired in the OF network. Therefore, online symmetric matrix factorization may serve as an algorithmic theory of neural computation.

## I. INTRODUCTION

In the quest to understand neural computation in mammals, the primary visual cortex (V1) has been an attractive and well-studied target system [1]. One of its major tasks is computing orientationally selective responses, or Gabor-filter receptive fields, out of orientationally nonselective inputs [2]. Such computation has been successfully modeled by Olshausen and Field (OF) who proposed a neural network that learns Gabor-filter receptive fields from an ensemble of natural images in an unsupervised fashion [3,4]. The OF network appeals as a model of V1 because it is both rigorously derived from a principled cost function and captures several salient anatomical and physiological features of V1 networks [5].

However, there remains an unanswered question regarding modeling neural computation in V1 by the OF algorithm. Whereas the original two-layer neural network implementation of the OF algorithm [3,4] may model sensory periphery [11,12] the required symmetric feedback connections have not been observed in V1. At the same time, in the single-layer network implementation of the OF algorithm [6] appropriate for V1, the learning rule derived from the OF cost function is nonlocal - the synaptic weight update depends on the activity of neurons other than just pre- and postsynaptic ones - and therefore biologically implausible.

In this paper, we propose a novel cost function and demonstrate that from it one can derive neuronal dynamics and local learning rules, both Hebbian for feedforward and anti-Hebbian for lateral synaptic connections. We demonstrate that training the network on a natural image ensemble yields Gabor-filter receptive fields. We also demonstrate that the application of such rules yields lateral connection weights that obey the same relationship with feedforward weights as in the OF framework. In addition, our framework accounts for several salient properties of biological networks and predicts that the learning rate decays with time in an activity-dependent fashion agreeing with experiments. Therefore, we make a step towards understanding V1 and mammalian neural computation in general.

The paper is organized as follows. In the next Section we summarize the OF algorithm and its neural network implementation. In the Results: A) we present the new cost function, the regularized error squared between the input's and the output's similarity matrices, and a derivation of an online algorithm for sparse dictionary learning with local learning rules; B) we report the results of numerical simulations showing that our network performs similarly to OF; C) we derive analytically the observed relationship between feedforward and lateral synaptic connection weights in our network, which reproduces a hard-wired constraint in the OF network; D) we show that our online symmetric matrix factorization algorithm can discover independent components in the whitened input data. In the Discussion: A) we compare our model to biology; B) we suggest that matrix factorization may be a generic model of neural computation.

## II. THE OLSHAUSEN-FIELD (OF) ALGORITHM

To motivate our work we briefly review the OF model [3,4] and point out the biologically implausible aspect of the single-layer implementation. The starting point of the OF model is the assumption that the vectorized image patches, $x_t \in R^n$, are represented by the neuronal feature vectors, i.e. columns of an overcomplete ($m>n$) dictionary, $W \in R^{n \times m}$, weighted by a sparse vector of neuronal activities, $y_t \in R^m$. To obtain such representation the OF model minimizes the squared representation error regularized by the $l_1$-norm of activity:

$$\min_W \sum_t \min_{y_t} \left( \frac{1}{2} \|x_t - W y_t\|_2^2 + \lambda \|y_t\|_1 \right), \quad (1)$$

where $\lambda$ reflects the relative importance of sparsity and representation accuracy.

To derive a neural network algorithm, OF minimized (1) in response to sequentially presented natural image patches, a so-called online setting. Specifically, for each presented image, $x_t$, they i) find the optimal value of $y_t$ for fixed $W$, ii) for fixed $y_t$ perform stochastic gradient descent with respect to feature vectors, $W$. Next, we discuss these two steps in more detail.

i) To find $y_t$ for each image the algorithm minimizes (1) using stochastic (sub)gradient descent steps [7] with respect to $y_t$:

$$\begin{cases} c_t = W'_t x_t - W'_t W_t y_{t-1} \\ y_t = \mathrm{ST}(c_t, \lambda) \end{cases}, \quad (2)$$

where ST is a component-wise soft-threshold function [8], see Fig. 1A. Equation (2) can be viewed as dynamics of activity in a single-layer network with feedforward and lateral connections [6], Fig. 1B. Then, $c_t$ represents the total input currents into neurons and soft thresholding models a rectifying nonlinearity of a biological neuron. In such network, lateral connections implement "explaining away", or competition between neurons in representing an input signal.

ii) After the network activity $y_t$ converges to a representation of an image, the algorithm updates feature vectors, $W$:

$$W_{t+1,i,j} = W_{t,i,j} + \delta \left( x_{t,i} - \sum_k W_{t,i,k} y_{t,k} \right) y_{t,j}, \quad (3)$$

where $W_{t+1,i,j}$ can be viewed as the synaptic weight for the connection from neuron $j$ to $i$, $\delta > 0$ is the learning rate.

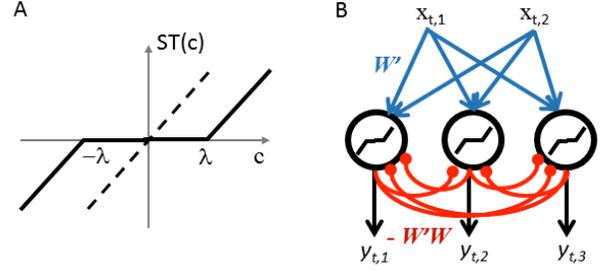

Fig. 1: A neural network implementation of the OF algorithm. A) Soft Thresholding (ST) function. B) A single-layer OF network. Each neuron applies ST on the inputs weighted by the feedforward connections, $W'$, minus outputs weighted by the lateral connections, $W'W$. Connection weights are updated using nonlocal learning rules.

The OF model (2,3) successfully reproduces several salient features of the primary visual cortex (V1) anatomy and physiology such as the overcompleteness of cortical representation, sparsity of neural activity, nonlinearity of neural responses [4,5]. Perhaps, most impressively, the receptive fields (computed from the feature vectors and whitening matrix) learned by the network on the ensemble of whitened natural images are Gabor-filter patches resembling receptive fields of neurons in V1 [3,4].

However, a major problem with modeling V1 with the OF algorithm is that in the single-layer network implementation [6] the learning rules are nonlocal. Specifically, the proposed learning rule (3) requires that each synapse "knows" the weights of synapses belonging to neurons other than its pre- and postsynaptic neuron. Because no mechanism exists for such communication in the brain it is not clear how the OF model can describe learning in V1. In addition, lateral connection weights in the OF model (2) are not learned directly but computed from the feedforward connection weights, i.e. the lateral connection matrix $M$ satisfies

$$M = -W'W . \quad (4)$$

Previously, this problem was addressed by a network of OF architecture but with local learning rules: Hebbian for feedfoward and anti-Hebbian for lateral connections [9,10]. However, such local learning rules have been postulated rather than derived from any cost function.

## III. RESULTS

Here, we derive a single-layer network for sparse overcomplete representation by minimizing the cost function comprising the squared difference between the similarity matrices of the input and the output data and a sparsity-inducing regularizer. Next, we demonstrate that this network learns Gabor patch receptive fields when trained on a natural image ensemble. Furthermore, we show that the relationship between lateral and feedforward connection weights agrees with that hard-wired into the OF network. Interestingly, our framework also predicts the decay of learning rate with time as observed experimentally.

## A. Cost function and derivation of the algorithm

We start by introducing a data matrix notation for algorithm input:

$$X = (x_1 \cdots x_T) = \begin{pmatrix} x_{1,1} & \cdots & x_{T,1} \\ \vdots & \ddots & \vdots \\ x_{1,n} & \cdots & x_{T,n} \end{pmatrix}, \quad (5)$$

and for algorithm output:

$$Y = (y_1 \cdots y_T) = \begin{pmatrix} y_{1,1} & \cdots & y_{T,1} \\ \vdots & \ddots & \vdots \\ y_{1,m} & \cdots & y_{T,m} \end{pmatrix} = \begin{pmatrix} y_{\bullet,1} \\ \vdots \\ y_{\bullet,m} \end{pmatrix}. \quad (6)$$

We denote a transpose of matrix $A$ as $A'$ and its Frobenius norm as $\|A\|_F$.

We propose to model the online sparse dictionary learning by minimizing the following cost function:

$$y_T = \arg\min_{y_T} \|X'X - Y'Y\|_F^2 + \lambda \sum_i \|y'_{\bullet,i} y_{\bullet,i}\|_1 \quad (7)$$

where $y_{\bullet,i}$ is an $i$-th row of matrix $Y$ (6), the activity of $i$-th output channel. The same loss term without the regularizer has been used previously, in the offline setting, in multi-dimensional scaling [13] and in symmetric nonnegative matrix factorization, where $Y$ is constrained to be element-wise nonnegative [14]. Whereas the regularizer may not look familiar, it induces sparsity on the outer product of rows of $Y$ and hence the activity of output channels. The motivation for choosing the particular form of the sparsity inducing regularizer will become clear below.

Let us derive an online algorithm temporarily ignoring the regularizer in (7). Such minimization problem can be solved by taking a derivative with respect to $Y$ and setting it to zero:

$$\left[ \frac{\partial}{\partial Y} \|X'X - Y'Y\|_F^2 \right]_{T,\bullet} =$$

$$= \left[ \frac{\partial}{\partial Y} \text{Tr}(Y'YY'Y - 2X'XY'Y) \right]_{T,\bullet} = \quad (8)$$

$$= \left[ 4(YY'Y - YX'X) \right]_{T,\bullet} = 0,$$

where the subscript $T,\bullet$ denotes the $T$-th column. When $T > m$, the products $YY'$ and $YX'$ change slowly with time and can be approximated by the matrices $\tilde{M}_T$ and $\tilde{W}_T'$ computed on the data available before the presentation of $T$-th sample. Then (8) can be linearized:

$$\tilde{M}_T y_T = \tilde{W}_T' x_T, \quad (9)$$

This linear system can be solved by coordinate descent (to avoid matrix division) leading to the following dynamics of neuronal activity:

$$y_{T,i} \leftarrow W_{T,i}' x_T - M_{T,i} y_T,$$

where

$$W_{T,j,i} = \frac{\sum_{t=1}^{T-1} y_{t,i} x_{t,j}}{\sum_{t=1}^{T-1} y_{t,i}^2}; \quad M_{T,i,j \neq i} = \frac{\sum_{t=1}^{T-1} y_{t,i} y_{t,j}}{\sum_{t=1}^{T-1} y_{t,i}^2}; \quad M_{T,i,i} = 0. \quad (10)$$

These expressions lead to a natural single-layer network implementation of the algorithm, Fig 2A, where matrices $W$ and $M$ correspond to feedforward and lateral synaptic connection weights correspondingly. Interestingly, although the synaptic weights did not appear explicitly in the cost function (7), they arise naturally in the online minimization algorithm (10).

Importantly, unlike in the single-layer neural network implementation of the OF model (3), here, the expressions for the synaptic weights are local, i.e. depend on the activities of only pre- and postsynaptic neurons, Fig 2A.

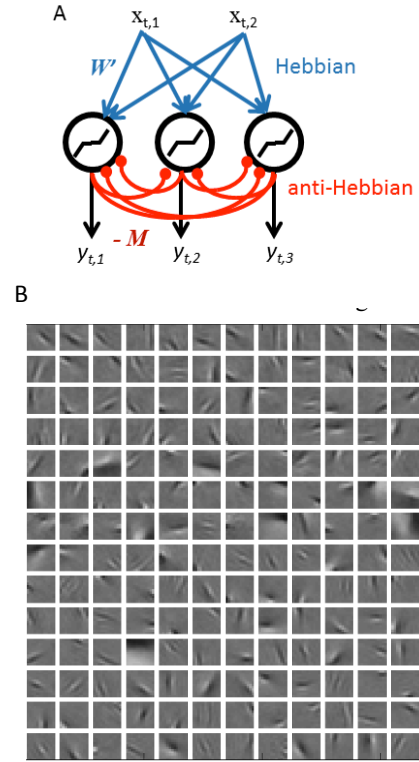

Fig. 2: A neural network implementation of the sparse matrix factorization algorithm. A) A single-layer network with local learning rules. Each neuron applies Soft Thresholding (ST) to the inputs weighted by the feedforward connections, $W'$, minus outputs weighted by the lateral connections, $M$. Connection weights are updated using Hebbian and anti-Hebbian learning rules correspondingly. B) Receptive fields learned on the whitened natural image ensemble.

To avoid storing past input and output activity appearing in the sums (10) we rewrite learning rules in a recursive form that admits online implementation:

$$\hat{Y}_{T,i} = \hat{Y}_{T-1,i} + y_{T-1,i}^2,$$
$$W_{T,j,i} = W_{T-1,j,i} + y_{T-1,i}\left(x_{T-1,j} - W_{T-1,j,i}y_{T-1,i}\right)/\hat{Y}_{T,i}, \quad (11)$$
$$M_{T,i,j\neq i} = M_{T-1,i,j} + y_{T-1,i}\left(y_{T-1,j} - M_{T-1,i,j}y_{T-1,i}\right)/\hat{Y}_{T,i}.$$

Thus, the feedforward synaptic weights are updated according to the Oja's modification of the Hebb rule [15] with the activity dependent learning rate. To the best of our knowledge such single-neuron learning rule [16] has not been previously derived for the multi-neuron case. Moreover, for the first time, we were able to derive the Oja-like version of the anti-Hebbian rule, see also [17,18].

Including the regularizer in the cost function alters the derivation in that instead of the derivative one needs to take a sub-derivative [7]. This does not affect the learning rules but adds soft thresholding [8] of the inputs to the dynamics:

$$y_{T,i} \leftarrow \mathrm{ST}\left(W_{T,i}'x_T - M_{T,i}y_T, \eta_{T,i}\right), \quad (12)$$

where the threshold is:

$$\eta_{T,i} \equiv \frac{\lambda}{2}\frac{\sum_{t=1}^{T-1}|y_{t,i}|}{\sum_{t=1}^{T-1}y_{t,i}^2}. \quad (13)$$

Now, our motivation for the choice of the regularizer in (7) should become clear: the regularizer was chosen in order to preserve the magnitude of the threshold with time. When output activity is binary, 0 or 1, as in a spiking neuron, the threshold stays exactly the same. When output activity is real, corresponding to the firing rate model or graded potential neurons, the constancy is only approximate but has been confirmed by numerical simulations.

Thus, we derived an online algorithm that can be implemented by a single-layer network with OF architecture relying only on local learning rules. Next we simulate our algorithm numerically by training it on the ensemble of natural images.

*B. Numerical simulations*

We applied our algorithm (11,12) to a natural image ensemble. Specifically, $10^4$ 12×12 pixel patches randomly extracted from natural images [19] and whitened. The extracted principal components were presented sequentially to a network of 196 neuron with feedforward and lateral connections, Fig. 2A. While each patch was presented, the coordinate descent update (12) was repeated 50 times for each neuron. Therefore, we simulate the neural dynamics with a total of $5\times10^5$ iterations. We initialize the network connection weights with Gaussian random variables and output activity with zeros. We set the initial synaptic learning rate to be $1/\hat{Y}_{1,i} = 10^{-4}$ and the initial firing threshold to be $\eta_{1,i} = 1.0$.

As a result of training, the network learns the feedforward weight matrix, *W'*. To plot neural filters (or receptive fileds) acting on natural image patches, we right-multiply *W'* by the whitening matrix *Q* and plot the rows, Fig. 2B. One can see that the receptive fields have the appearance of Gabors filters of varying orientation and spatial frequency. We fit the receptive fields with 2D Gabor functions:

$$G(\tilde{x},\tilde{y}) = g\exp\left(-\tilde{x}^2/2\sigma_{\tilde{x}}^2 - \tilde{y}^2/2\sigma_{\tilde{y}}^2\right)\cos\left(2\pi f\tilde{x} + \varphi\right),$$

where $\tilde{x} = (x-x_0)\cos\theta + (y-y_0)\sin\theta$ and $\tilde{y} = -(x-x_0)\sin\theta + (y-y_0)\cos\theta$ are obtained by a translation of the original coordinate system $(x_0, y_0)$ followed by a rotation by angle $\theta$. In this equation, $g$ is the amplitude, $\sigma_{\tilde{x}}$ and $\sigma_{\tilde{y}}$ represent the widths of the Gaussian envelope, $f$ is the spatial frequency of the sinusoidal grating, and $\varphi$ is phase offset. We present the measured distribution of spatial frequencies and orientation in Fig. 3A and B respectively. Both statistics were similar to that in the OF network [19] and in physiological measurements [28]. Furthermore, the distribution of output activity, *y*, has a strong peak at zero (sparsity) and a heavy tail, Fig. 3C.

Finally, we found that the feedforward and lateral connection weights are strongly correlated, Fig. 4. Whereas in the OF network such correlation, (4), is predetermined by the algorithm (2,3), in our network it appeared as a result of independently acting learning rules.

*C. Derivation of the relationship between feedforward and lateral connections*

In this Section we present an analytical derivation of the relationship between connection matrices *W* and *M* in the steady state solution of the sparse symmetric matrix factorization cost function. Because the dictionary is overcomplete, when the regularization constant, $\lambda$, is not too large, the steady state solution satisfies approximately:

$$X'X = Y'Y \quad (14)$$

The SVD of the data matrix *X* can be written in a standard form:

$$X = U_X\Sigma_X V_X', \quad (15)$$

where as usual singular vectors are orthonormal, $U_X'U_X = I$, $V_X'V_X = I$ and $\Sigma_X$ is a diagonal matrix. Similarly,

$$Y = U_Y \Sigma_Y V_Y', \qquad (16)$$

Next, we substitute (15,16) into (14):

$$V_X \Sigma_X' U_X' U_X \Sigma_X V_X' = V_Y \Sigma_Y' U_Y' U_Y \Sigma_Y V_Y'.$$

By taking into account the orthonormality of the singular vectors:

$$V_X \Sigma_X' \Sigma_X V_X' = V_Y \Sigma_Y' \Sigma_Y V_Y'$$

From this we conclude that the right singular vectors of $X$ and $Y$ are equal, and $\Sigma_X$ and $\Sigma_Y$ share the same nonzero diagonal values:

$$V_X = V_Y = V, \text{ and } \Sigma_Y' \Sigma_Y = \Sigma_X' \Sigma_X. \qquad (17)$$

Then, the unnormalized connection weight matrices for feedforward and lateral connections:

$$\tilde{W}' = YX' = U_Y \Sigma_Y V' V \Sigma_X' U_X' = U_Y \Sigma_Y \Sigma_X' U_X', \qquad (18)$$

$$\tilde{M} = YY' = U_Y \Sigma_Y V' V \Sigma_Y' U_Y' = U_Y \Sigma_Y \Sigma_Y' U_Y'. \qquad (19)$$

Note that

$$\begin{aligned}\tilde{W}'\tilde{W} &= U_Y \Sigma_Y \Sigma_X' \Sigma_X \Sigma_Y' U_Y' \\ &= U_Y \Sigma_Y \Sigma_Y' \Sigma_Y \Sigma_Y' U_Y' = U_Y \left( \Sigma_Y \Sigma_Y' \right)^2 U_Y'.\end{aligned} \qquad (20)$$

If the input matrix is properly whitened $\Sigma_x$ contains only 1's or 0's on the diagonal. Since $\Sigma_x$ and $\Sigma_y$ share the same nonzero diagonal values, $\Sigma_y$ also contains only 1's or 0's on the diagonal. Therefore, (19) and (20) are identical establishing a relationship between feedforward and lateral connection weights.

To obtain synaptic connection weights, $W'$ and $M$, one has to normalize $\tilde{W}'$ and $\tilde{M}$ by the cumulative postsynaptic

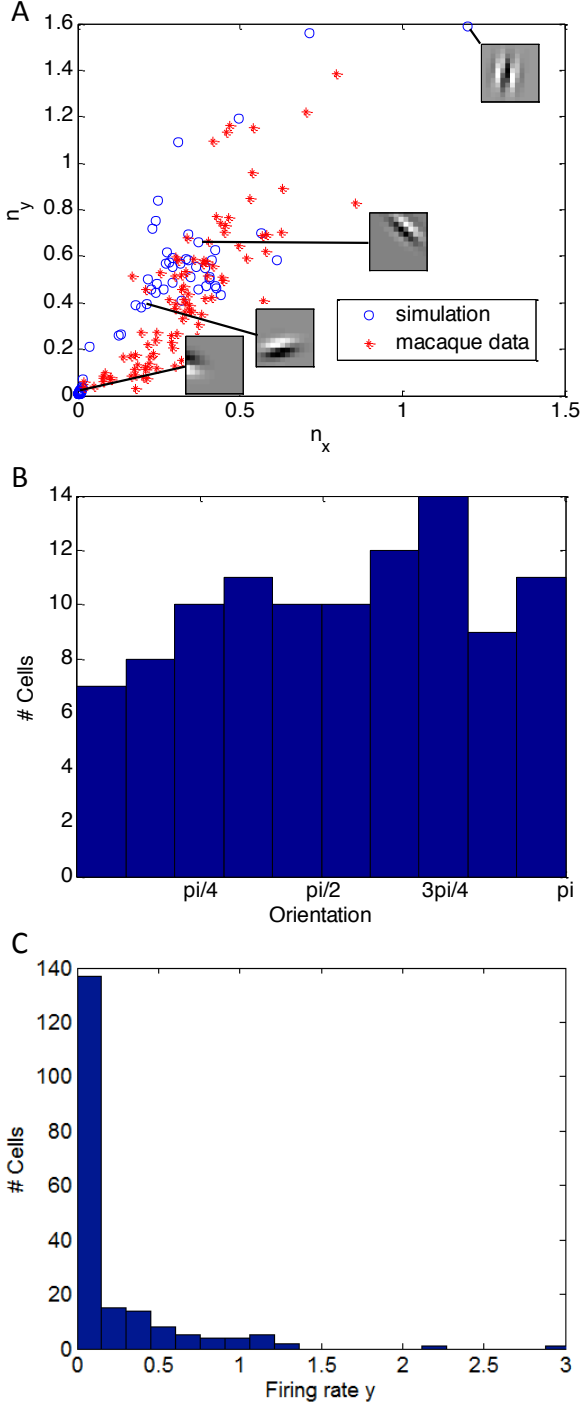

Fig. 3: Statistics of receptive fields and neuronal activity computed in our network matches that of OF model [19] and mammalian physiology [20, 28]. A. Spatial frequencies of Gabor fits, where $n_x = f\sigma_x$ and $n_y = f\sigma_y$. B. The distribution of orientation preference, $\theta$ C. The distribution of activity, $y$, among output units is sparse and heavy-tailed.

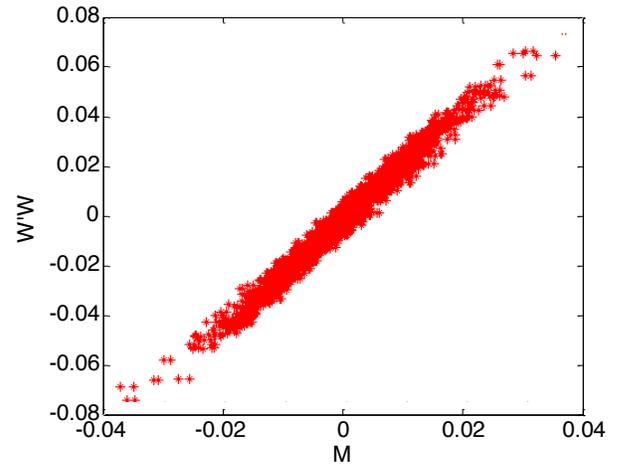

Fig. 4: Correlation between the lateral connection weights and the Gramm matrix of the feedforward connections (off-diagonal elements only) in the sparse matrix factorization network.

activity (10). The normalization accounts for the variation in slope in Fig.4.

*D. Sparse symmetric matrix factorization can disover independent components*

Here we argue that symmetric matrix factorization can be used to discover independent components in their whitened mixture, i.e. perform independent component analysis (ICA). ICA has been successful in recovering Gabor filters from natural images [22]. The argument given below can be seen as an alternative explanation of our numerical simulation results given in Section III B.

The goal of ICA is to recover the sources, given that the input data are generated by the following linear model [21]:

$$x_T = As_T, \qquad (21)$$

where $A \in R^{n \times n}$ is the mixing matrix, assumed to be invertible, and the random source vector, $s_T$ has statistically independent elements. Each source is assumed to have zero mean and sparse, e.g. Laplace distributed.

To establish a connection between sparse symmetric matrix factorization and ICA, we first show that the whitened input (21) is an orthogonal rotation of the original sources [19]. To see this, we rewrite the whitened input, $\tilde{x}_T$ in terms of the assumed to be known whitening matrix, $Q$, and by substituting (21) find:

$$\tilde{x}_T = Qx_T = QAs_T. \qquad (22)$$

By denoting $QA \equiv G$ we obtain from (22):

$$\tilde{x}_T = Gs_T.$$

The orthonormality of the square matrix $G$ follows from the ortonormality of whitened data [19]:

$$I_n = \langle \tilde{x}_T \tilde{x}_T' \rangle_T = G \langle s_T s_T' \rangle_T G' = GG'. \qquad (23)$$

To demonstrate that our algorithm can be used as online ICA, we rewrite the cost function (7) for the whitened input by using the orthonormality $G$ (23):

$$y_T = \arg\min_{y_T} \left\| \tilde{X}'\tilde{X} - Y'Y \right\|_F^2 + \lambda \sum_i \left\| y'_{\cdot,i} y_{\cdot,i} \right\|_1$$
$$= \arg\min_{y_T} \left\| S'S - Y'Y \right\|_F^2 + \lambda \sum_i \left\| y'_{\cdot,i} y_{\cdot,i} \right\|_1. \qquad (24)$$

Because $Y$ has the same dimensionality as $S$, $Y = S$ is a minimum of the unregularized cost in (24). However, this minimum is not unique: $Y$ can be left-rotated by an orthonormal matrix without affecting the cost. Then, the role of the sparsity-inducing regularizer is to favor a sparse $Y$, allowing the recovery of the original sparse $S$.

The analysis of this section can be extended straightforwardly to the offline ICA problem. Symmetric matrix factorization or whitened input, with a suitably chosen sparsity inducing regularizer can be used as an ICA cost function.

## IV. DISCUSSION

In this paper, by introducing a novel cost-function we derived an online algorithm that reproduces many features of the OF model but can be implemented by a single-layer neural network relying only on local learning rules. Therefore, we proposed a more biologically plausible implementation of the sparse coding hypothesis.

*A. Biological relevance*

1. Weighted summation of inputs and soft thresholding. Our online algorithm maps onto a neural network where each unit performs soft thresholding of the weighted sum of its inputs (both feedforward and lateral). Such computation corresponds to a commonly used basic model of biological neurons. Although, the two-sided thresholding our algorithm requires is not encountered in biological neurons, it may be implemented by a pair of neurons each responsible for positive or negative inputs. Such ON and OFF neurons exist in the peripheral visual system of both vertebrates and inveretebrates [27].

2. Local Hebbian and anti-Hebbian synaptic learning rules. The learning rules we derived are consistent with those previously abstracted from biological observations of synaptic plasticity. Crucially, these learning rules do not require any synapse to keep track of the activity of neurons other than the pre- and postsynaptic pair it connects. Anti-Hebbian learning could be implemented indirectly via a Hebbian update of the synaptic weights of inhibitory interneurons.

3. Dependence of learning rate on cumulative activity. The learning rate in the synaptic weight update is inversely proportional to the cumulative activity of the postsynaptic neuron (11). Such variation of plasticity with time corresponds to the reports of LTP decaying with age in an activity dependent manner [23-25].

4. Sparsity of neuronal activity. The distribution of neuronal firing has a peak at zero and a heavy tail, Fig. 3C in agreement with physiological measurements [16,26].

*B. Symmetric matrix factorization as a generic model of neural computation*

We believe that the significance of online symmetric matrix factorization goes beyond deriving sparse dictionary learning with local Hebbian and anti-Hebbian learning rules. We speculate that it serves as a powerful and versatile elementary building block of neural computation. Indeed, symmetric matrix factorization with (and without) various constraints can solve multiple computational objectives. We argued above that ICA can be formulated as a symmetric matrix factorization problem. Furthermore, unconstrained symmetric matrix factorization can compute the principal subspace of the streamed data [17]. Nonnegative symmetric matrix factorization can be viewed as a clustering algorithm capable of nonlinear feature discovery [18]. Jointly, these tools represent a formidable arsenal for modeling neural computation.

ACKNOWLEDGMENTS

The authors would like to thank Sanjeev Arora, Alex Genkin, Bruno Olshausen, Eftychios Pnevmatikakis, Christopher Rozell, and Zaid Towfic for helpful discussions.